\documentstyle[preprint,aps]{revtex}
\begin{document}
\draft
\title{Quantum vortices near the superconductor-insulator transition
	in Josephson junction arrays}
\author{Anne van Otterlo$^{a)}$ and Rosario Fazio$^{b)}$}
\address{a)Institut f\"ur Theoretische Festk\"orperphysik,
	Universit\"at Karlsruhe,\\
	Postfach 6980, 76128 Karlsruhe, FRG\\
	b)Istituto di Fisica, Universit\`a di Catania,
	viale A. Doria 6, 95129 Catania, Italy}
\maketitle
\begin{abstract} The properties of vortices in Josephson junction arrays are
investigated in the quantum regime near the superconductor-insulator
transition.
We derive and study an effective action for vortex dynamics that is valid in
the region
where the charging energy is comparable to the Josephson coupling energy.
In the superconducting phase the onset of quantum effects
reduces the vortex mass and depinning current.
In the case of long range Coulomb interaction between Cooper pairs we find that
as the
transition is approached, the velocity window in which ballistic vortex motion
is possible grows.
At the superconductor-insulator transition the vortex mass vanishes
and vortices and spinwaves decouple.
In the case of on-site Coulomb repulsion (which is of
relevance for superconducting granular films) the vortex mass
it is sample-size dependent in the superconducting phase, but
stays finite at the critical point where it is scale invariant.
The relation of our work to experiment is discussed.
\end{abstract}
\pacs{PACS numbers: 74.50, 74.60.Ge, 74.65+n}

\narrowtext

\section{Introduction}

In classical Josephson junction arrays (JJA) vortices are well known
topological excitations that characterize both the dynamical and
thermodynamical properties of these systems. At low temperatures these
excitations
are bound in dipoles of opposite vorticity and at a critical temperature they
unbind
in a Kosterlitz-Thouless-Berezinskii (KTB) phase transition \cite{kn:KTB}
leading
the system from the superconducting (SC) to a resistive phase.

When an external driving current is applied to the array, free vortices (as
induced by
applying a magnetic field) move and
a voltage drop accross the sample appears. In classical JJA's, in
which the Josephson coupling $E_{J}$ is much larger than
the charging energy $E_{C}$, vortex motion is diffusive. Some time ago it was
realized
that charging effects yield dynamics and a mass for
the vortices in the SC phase \cite{kn:sim,kn:ueas,kn:foszm}.
The electrostatic energy stored in the junction capacitances
can be interpreted as a kinetic energy due to the vortex motion and
as a consequence a mass can be attributed to the vortex. The mass in this
regime was found to be
proportional to the junction capacitance C and in case of small damping
ballistic
vortex motion was predicted. Modern lithographic techniques allow
for the realization of JJA's with junction resistance much larger than
$h/e^{2}$, thereby making possible an experimental check of these predictions.
Very recently, in an important experiment, van der Zant et al.
\cite{kn:herb} demonstrated the existence of such ballistic vortices in
triangular arrays.

The possibility of ballistic motion depends
crucially on the dissipation mechanisms available to a vortex.
Nowadays it is possible to fabricate
high quality JJA's in which no ohmic dissipation is present \cite{kn:herc}.
If the vortex velocities are below a certain threshold
(related to the superconducting gap), also the
quasiparticles are frozen and do not contribute to the damping.
It was pointed out in recent analytical \cite{kn:ulig,kn:ulie}
and numerical \cite{kn:pebo} studies that another mechanism by which
a classical moving vortex can loose its
kinetic energy is by emitting spinwaves. Depending on the spinwave
spectrum, it has been shown that in triangular arrays there exists a small
window
of velocities for the vortex to move over the pinning potential provided by the
lattice without suffering too much damping. For square classical arrays
ballistic
vortex motion is not possible.

All the previous investigations were mainly concerned with the case in
which the Josephson coupling $E_{J}$ is much larger than
the charging energy $E_{C}$ (and larger than the temperature $T$).
When these two energies become comparable quantum effects play a major role
and,
eventually, at a certain critical value of their ratio,
a superconductor-insulator (S-I) transition at
zero temperature takes place \cite{kn:SI,kn:doni,kn:sage}.
This phase transition separates regions where either the Cooper pairs or the
vortices are localized. In the insulating phase vortices are
delocalized, as are the Cooper pairs in the SC phase.
Thus, close to the S-I transition one expects vortex properties to
reflect the presence and nature of the transition. One may also
expect that experiments on vortices in this region \cite{kn:Tink} yield
more insight into the phase transition.

The analysis of vortex dynamics as one approaches the S-I transition will be
the aim of
this paper. We will consider a square JJA at $T=0$ with a superimposed
external magnetic field such that a small amount of
vortices of one sign are induced in the system. If the magnetic frustration is
very small
one may suppose that the vortex motion is not influenced by their mutual
interaction.
By applying an external current the dynamics can be studied. When $E_{J}\gg
E_{C}$
the description has been formulated in
\cite{kn:sim,kn:ueas,kn:foszm,kn:ulig,kn:ulie}.
Vortex motion is then described by a phenomenological equation of motion for
the vortex coordinate $x$ of the form \cite{kn:orla}
\begin{equation}
	M_{v}\ddot{x}+\eta \dot{x}=2\pi E_{J} I/I_{cr} +
	\pi U_{bar} \sin(2\pi x) \; ,
\label{eq:phen}
\end{equation}
where $M_{v}$ is the vortex mass, $\eta$ a phenomenological damping, $I$ the
applied
current, $I_{cr}$ the junction critical current, and $U_{bar}$ the height of
the
potential energy barrier for a vortex to go from one plaquette to a neighboring
one
in the potential landscape provided by the array.
When we enter the quantum regime fluctuations of vortex anti-vortex
pairs (dipoles) start to be relevant and they will interact with the moving
vortex.
As a result the vortex dynamics will be modified.

In order to describe this regime, we first derive an effective action
for one single vortex in the next section. We will show that it depends on
the charge-charge correlation function; Eq.(\ref{eq:effect}) is the central
result of our work. It implies that a description in terms of an equation of
motion is still possible, however, with modified coefficients.
General expressions for the vortex mass and the spin wave
damping that are not constrained by the inequality $E_{J}\gg E_{C}$ and are
also valid when the Josephson energy is comparable
with the charging energy are then presented. After that we briefly show
in Sect.\ref{sec-cla} how to recover
the previously known results in the limit of large Josephson coupling.
Analytic approaches in the two opposite limits in which either
the junction capacitance or the capacitance to the ground is dominant are
discussed
in Sects.\ref{sec-sca} and \ref{sec-cog} respectively. For long range
Coulomb interactions we find that closer to the S-I transition the vortex mass
clearly deviates from the classical results and is not simply proportional
to the junction capacitance. Furthermore a velocity window for ballistic motion
opens
if one approaches the transition, also for square arrays in which ballistic
motion
in the limit $E_{J}\gg E_{C}$ is impossible.
The effect of the capacitance to the ground has until now been overlooked for
JJA's,
but it gives a contribution to the vortex mass which is of the same order of
magnitude (for a typical experimental setup) as the mass due to
the junction capacitance and therefore, we argue, it should be taken into
account.
In the limit of short range Coulomb interactions we notice that
the mass should be finite at the transition and the
vortex dynamics is governed by the same correlation functions that lead
to the universal conductance at the S-I transition. Therefore we feel that
the understanding of vortex dynamics in this regime may shed light on the
source
of dissipation which is responsible for the metallic behavior at the S-I
transition
\cite{kn:anne,kn:self}.
Sect.\ref{sec-mcs} discusses Monte Carlo simulations that verify our
conclusions.
The main results of our work are then summarized in the last section.

\section{The effective action}
\label{sec-eff}

We start from the well known Hamiltonian for a Josephson junction array
\begin{equation}
	H=\frac{1}{2} \sum_{ij} Q_{i} C^{-1}_{ij} Q_{j} -
	E_{J} \sum_{<ij>} \cos(\phi_{i}-\phi_{j})+
	\sum_{i}\vec{I}_{i}\cdot\vec{\nabla}\phi_{i} \; ,
\label{eq:hhh}
\end{equation}
where $Q_{i}$ and $\phi_{i}$ denote respectively the charge and the phase
of the superconducting order parameter of the i-th island. The external current
is denoted by $I$. The typical energy scales are the Josephson
coupling $E_{J}$ and the charging energy $E_{C}= e^{2}/2C$ or
$E_{o}=e^{2}/2C_{o}$.
$C_{o}$ and $C$ are, respectively, the ground capacitance
and the nearest neighbor capacitance that constitute the capacitance matrix
$C_{ij}$.
The range of the electrostatic interaction between Cooper pairs, which is
described
by the inverse capacitance matrix $C^{-1}$, is $\lambda^{-1}=\sqrt{C/C_{o}}$.
The quantum mechanical description is completed by the commutation relation
$[Q_{i},\phi_{j}]=2e\delta_{ij}i$.

By means of well established duality transformations \cite{kn:Savit} it is
possible to recast the partition function in terms of the topological
excitations of the system. For the Hamiltonian under consideration this was
done
in Refs.\cite{kn:sage,kn:fgs}, where a detailed derivation
is presented. In order to make this paper self-contained we sketch
the main steps in appendix \ref{app-dtr}. The result is a description in terms
of
two discrete gasses $q$ and $v$ that denote charges and vortices respectively.
In this formulation the partition function for a JJA is obtained summing over
all configurations of charges and vortices \cite{kn:sage}
$$
	Z=\sum_{\{q_{i,\tau}\}}\sum_{\{v_{i,\tau}\}} \exp\{-S[q,v]\}
$$
where the action is
$$
	S[q,v]= \epsilon \pi E_{J}\sum_{ij,\tau} v_{i,\tau}G_{ij}v_{j,\tau}
	+\frac{2\epsilon E_{C}}{\pi} \sum_{ij,\tau} q_{i,\tau}U_{ij}q_{j,\tau}
$$
\begin{equation}
	+i\sum_{ij,\tau} \dot{q}_{i,\tau}\Theta_{ij}v_{j,\tau}
	+i\epsilon\sum_{ij,\tau}
\vec{I}_{i,\tau}\cdot\vec{\nabla}\Theta_{ij}v_{j,\tau}
	+\frac{1}{4\pi \epsilon E_{J}} \sum_{ij,\tau}
	\dot{q}_{i,\tau}G_{ij}\dot{q}_{j,\tau}
\label{eq:axi}
\end{equation}
where the kernels
$\Theta_{ij}=\arctan\left(\frac{y_{i}-y_{j}}{x_{i}-x_{j}}\right)$
and $G_{ij}=-\ln\!\mid\! r_{i}-r_{j}\!\mid$ were introduced. $U_{ij}$ is
related to the
capacitance matrix as $U_{ij}= 2\pi C C^{-1}_{ij}$, it has the explicit form
$U_{ij}=K_{0}(\lambda\mid r_{i}-r_{j}\mid)$, where $K_{0}$ is a modified Bessel
function \cite{kn:sage}. Spinwaves are described
by the last term in Eq.(\ref{eq:axi}). We introduced $N_{\tau}$ time-slices to
describe
the quantum dynamics of the system. The lattice spacing
in the time direction is denoted by $\epsilon$ and
$N_{\tau}\epsilon=\beta$ is the inverse temperature (see App.\ref{app-dtr}).
The phase diagram for the model described by (\ref{eq:axi})
was studied in ref.\cite{kn:sage} and exhibits a zero temperature
S-I transition (an analysis based on a Kosterlitz type RG procedure is
in progress \cite{kn:RGanne}).

Eq.(\ref{eq:axi}) will be our starting point. We seek an effective
action for a single vortex (with coordinate $r(\tau)$) that includes the effect
of
the interaction
with fluctuating charges and other vortices (present in the system because of
quantum fluctuations). It is obtained by a summation in the
partition function
over all configurations of the charges and other vortices. It turns out to be
more
transparent to introduce the trajectory of the vortex
$v_{i,\tau}=v\delta(r_{i}-r(\tau))$. Here $v=\pm 1$ for a vortex or anti-vortex
respectively.
Formally the effective action can now be written as
\begin{eqnarray}
\nonumber
	S_{eff}= -\ln{\Big \langle} \epsilon 2\pi E_{J} v\sum_{ij,\tau}
	v_{i,\tau}G_{ij}\delta(r_{j}-r(\tau))+\\
	+iv\sum_{ij,\tau} \dot{q}_{i,\tau}\Theta_{ij}\delta(r_{j}-r(\tau))
	+iv\epsilon\sum_{ij,\tau}
\vec{I}_{i,\tau}\cdot\vec{\nabla}\Theta_{ij}\delta(r_{j}-r(\tau))
	{\Big \rangle}_{(\ref{eq:axi})} \;\; ,
\label{eq:1ststep}
\end{eqnarray}
where the average is to be taken with the action (\ref{eq:axi}). The first term
describes the static interaction with other vortices, whereas the second
describes
the dynamical interaction with charges.
This expression is formally exact, but difficult to evaluate because of
the nonlinearity of the action (\ref{eq:axi}). As we are interested
in the kinetic contribution (quadratic in the vortex velocities) it is,
however, sufficient to expand the average in
(\ref{eq:1ststep}) in cumulants and stop at the second order in the
vortex velocities $\dot{r}(\tau)$. For a uniform external current distribution
we find
$$
	S_{eff}=\frac{1}{2}\sum_{\tau\tau'} \dot{r}^{a}(\tau) {\cal M}_{ab}(r(\tau)-
	r(\tau'),\tau-\tau')\dot{r}^{b}(\tau') +
	2\pi i v\epsilon\sum_{\tau}\epsilon_{ab}I^{a}_{\tau}r^{b}(\tau)  \; ,
$$
\begin{equation}
	{\cal M}_{ab}=\sum_{jk} \nabla_{a}\Theta(r(\tau)-r_{j}) \langle q_{j\tau}
	q_{k\tau'}\rangle \nabla_{b}\Theta(r_{k}-r(\tau')) \; ,
\label{eq:effect}
\end{equation}
where $a,b=x,y$ and $\epsilon_{ab}$ is the anti-symmetric tensor.
Thus, {\it vortex dynamics is governed by the charge-charge
correlation}, which depends on the full
coupled charge vortex gas (CCVG) Eq.(\ref{eq:axi}). The effective action
Eq.(\ref{eq:effect}) describes the dynamical vortex properties for all values
of
$E_{J}/E_{C}$ and is therefore a good starting point for the investigation of
vortex properties near the S-I transition.

In order to obtain the vortex effective action in Eq.(\ref{eq:effect}) we
disregarded all higher order terms in the cumulant expansion.
This is certainly correct in the $E_{J}\gg E_{C}$ limit where (as
discussed in detail in the next section) the charges can be considered
as continuous variables and vortex fluctuations can be disregarded.
In general, however, the average defined in Eq.(\ref{eq:1ststep}) is far from
Gaussian ($q$ and $v$ are integer valued fields). As long as we
are in the superconducting phase, however, the charges are strongly fluctuating
and
the vortices are still bound in dipoles. Therefore we expect that the
higher order cumulants are still not important \cite{kn:Jose}.
A full description of the
vortex motion in the resistive region, nevertheless,
may require the analysis of a dynamical equation that contains
also terms proportional to higher powers of the velocity.

\section{the classical limit}
\label{sec-cla}

The expression (\ref{eq:effect}) yields the known results
\cite{kn:ueas,kn:ulig,kn:ulie} in the classical limit where $E_{J}\gg E_{C}$.
In this region of the phase diagram the Josephson coupling dominates over the
electrostatic energy. Far from the transition vortex fluctuations due to
quantum effects
are suppressed and they may be neglected. In this regime the charges are wildly
fluctuating (as to constitute the supercurrents that keep the SC phases well
defined)
and may be considered to be continuous variables.

Thus in the classical limit we may concentrate on the charge part of the action
(\ref{eq:axi}).
It is rewritten conveniently as
\begin{equation}
	S[q]=\sum_{ij\tau\tau'}q_{i\tau}Q_{ij\tau\tau'}q_{j\tau'} \;\; , \;\;\;
	Q_{ij\tau\tau'}=\frac{2\epsilon E_{C}}{\pi}\left( U_{ij}\delta_{\tau,\tau'}+
	\frac{G_{ij}}{\epsilon^{2}\omega^{2}_{p}}(2\delta_{\tau,\tau'}-
	\delta_{\tau,\tau'+\epsilon}-\delta_{\tau,\tau'-\epsilon})\right) \; ,
\label{eq:chke}
\end{equation}
where the plasma frequency $\omega_{p}=\sqrt{8E_{J}E_{C}}$ was introduced.
The charge-charge correlation is half of the inverse of the kernel $Q$.
It is
\begin{equation}
	\langle q q \rangle_{k,\omega_{\mu}}=
	(E_{J}k^{2}/\epsilon)/(\omega^{2}_{\mu}+\omega^{2}_{k}) \; , \;\;
	\omega^{2}_{k}=\omega^{2}_{p}U_{k}/G_{k} \; ,
\label{eq:clqq}
\end{equation}
The spinwave dispersion is described by $\omega_{k}$. It is optical,
i.e. $\omega_{k}=\omega_{p}$, for long range Coulomb interactions, whereas
for on-site interactions we have $\omega_{k}=\bar{\omega}_{p} k$.
Here $\bar{\omega}_{p}=\sqrt{8E_{J}E_{o}}$ is the plasma frequency
for the case of on-site Coulomb interactions.

The action (\ref{eq:effect}) reduces to that of a free particle in the
limit of small velocities $\dot{r}(\tau)$. Since the charge-charge correlation
(\ref{eq:clqq}) is short range in time we may put $r(\tau)=r(\tau')$ in
Eq.(\ref{eq:effect}).
The corresponding adiabatic vortex mass $M_{v}$ is
\begin{equation}
	M_{v}=\epsilon\sum_{\tau}{\cal M}_{xx}(0,\tau) \; ,
\label{eq:mfc}
\end{equation}
which reduces in the classical limit to $M_{v}= M_{ES}+\ln(L)\: M_{o}$,
where $M_{ES}=\pi^{2}/4E_{C}$ is the Eckern-Schmid mass \cite{kn:ueas} and
$M_{o}=\pi/8E_{o}$.
Thus {\it both} self- and nearest neighbour capacitances
yield a contribution to the mass. The self capacitance contribution
depends on the system size $L$. For generic sample
sizes and capacitance ratio's the new contribution (which has been
overlooked so far for JJA's since it makes no sense in the
thermodynamic limit, see ref.\cite{kn:lege} for size
dependent vortex masses in different systems) is somewhat
smaller than the Eckern-Schmid mass.

The instanton action $S_{inst}$, related to a hop from one plaquette to
a neighbouring one, determines tunnel rates and
the depinning current. As vortex trajectory we now take $\dot{v}_{i\tau}=
v_{i,\tau+\epsilon}-v_{i,\tau}=\delta_{\tau,t}
[\delta_{i,x+a}-\delta_{i,x}]$ for a hop from
$x,t\rightarrow x+a,t+\epsilon$. This may be inserted in the CCVG action
(\ref{eq:axi}) to find the general result
\begin{equation}
	S_{inst}=\frac{1}{2} {\cal M}_{xx}(0,0) .
\label{eq:sfc}
\end{equation}
In the classical limit we recover all known results \cite{kn:Kor}, i.e. for
general
capacitance matrix
\begin{equation}
	S_{inst}=\frac{\pi E_{J}}{4\omega_{p}}
\left[\sqrt{\pi}\sqrt{\lambda^{2}+4\pi}+
	\frac{\lambda^{2}}{2} \ln\left(\frac{2\sqrt{\pi}}{\lambda}+
	\sqrt{1+\frac{4\pi}{\lambda^{2}}}  \right) \right],
\label{eq:act}
\end{equation}
which reduces to $S_{inst}=\pi^{3/2}E_{J}/4\bar{\omega}_{p}$ and
$S_{inst}=\pi^{2}E_{J}/2\omega_{p}$ for $C=0$ and $C_{o}=0$ respectively. The
general form
of an instanton action in the WKB approximation, which is proportional to the
square root
of mass times barrier height, $S_{inst}\sim\sqrt{M_{v}U_{bar}}$, determines the
barrier
$U_{bar}$ for a vortex to hop. The depinning current $I_{dep}$ is half the
barrier height.
Since deep in the classical limit $U_{bar}=E_{J}/5$ \cite{kn:lobb}, we may
establish
$S_{inst}=\pi\sqrt{5M_{v}U_{bar}/2}$ for the $C_{o}=0$ case.

The spinwave damping that a moving vortex experiences may also be calculated
from (\ref{eq:effect}). Varying the vortex coordinate $r^{a}(\tau)$ in
Eq.(\ref{eq:effect}) yields the equation of motion
\begin{equation}
	2\pi i\epsilon_{ab} I^{b}/I_{cr}=\frac{1}{\epsilon} \partial_{\tau}
	\sum_{\tau'}{\cal M}_{ab}(r(\tau)-r(\tau'),\tau-\tau')\dot{r}^{b}(\tau')
\label{eq:eom}
\end{equation}
and its constant velocity
solutions in the presence of an external current determine
the nonlinear relation between driving current and vortex velocity, once
the charge-charge correlation is analytically continued
(i.e. sending $i\omega_{\nu}\rightarrow \omega +i \delta$)
to real frequencies \cite{kn:ulie}.
The relevant information is in the real part of Eq.(\ref{eq:eom}), which reads
in Fourier
components and for a constant vortex velocity $\dot{\vec{r}}(\tau)=(v,0)$
\begin{equation}
	I^{y}/I_{cr}= \frac{v}{4} \int d\omega\int^{+\pi}_{-\pi} d^{2}k
	\frac{k^{2}_{y}}{k^{2}}
	[\delta(\omega-\omega_{k})+\delta(\omega+\omega_{k})] \delta(\omega-vk_{x}) \;
{}.
\label{eq:reom}
\end{equation}
The delta functions express
the spinwave dispersion (from the analytic continuation of
the charge-charge correlation) and
the vortex dispersion respectively. The overlap integral determines
the amount of dissipation a moving vortex suffers from coupling to spinwaves.
If we adopt the smooth momentum integration cut-off
$\int d^{2}k \rightarrow 2\pi\int^{\infty}_{0} dk k\exp(-k/\sqrt{2\pi})$
that was introduced in Ref.\cite{kn:ueas}, we recover in the classical limit
the results of Refs.\cite{kn:ulig,kn:ulie}, see Fig.1 for the current vs.
velocity
relation for long range Coulomb interactions. Note that the minimum
velocity that a vortex needs to move over the pinning potential of the lattice
follows from the phenomenological equation of motion Eq.(\ref{eq:phen}) by
demanding
that $\frac{1}{2}M_{v}v^{2}\ge U_{bar}$ and
is about 0.14$\omega_{p}$ (see the dotted line in Fig.1)
which is also the velocity at which the spinwave damping
sets in. Therefore ballistic vortex motion is almost impossible in classical
square arrays. In triangular arrays, however, a similar analysis yields a
somewhat wider
velocity window \cite{kn:ulie}. In the next section
we show how the inclusion of quantum effects contributes to the opening
of a more robust velocity interval, also for square arrays, where ballistic
motion can be
observed.

\section{Long range Coulomb interactions}
\label{sec-sca}

When the ratio $E_{J}$/$E_{C}$ decreases the charge-charge correlation must be
calculated beyond the classical approximation. We first consider the case of
long range
Coulomb interactions between Cooper pairs. According to the arguments given
in Refs.\cite{kn:sage,kn:peyo}, the
zero temperature S-I phase transition is presumably of the KTB type, as is the
finite temperature
transition to the resistive phase. This means that no dimensional cross-over
takes place
at zero temperature, or in other words the dynamical critical exponent $z$
equals zero.

Thus we are led to conclude that the vortex fugacity in the superconducting
phase
scales to zero in the renormalization group sense also at zero temperature.
This means that
the charge-charge correlation function in the SC phase may still be evaluated
in the absence of vortex
fluctuations. Therefore we consider again only the charge part
Eq.(\ref{eq:chke}) of the action
Eq.(\ref{eq:axi}), but in contrast to the classical limit we now treat the
charges as discrete
variables.

The charge-charge correlation function may be rewritten as (see Appendix
\ref{ss-lr} for the derivation) the classical result minus a correction
\begin{equation}
	\langle q_{j\tau}q_{k\tau'}\rangle
	=\frac{1}{2} Q^{-1}_{jk\tau\tau'} -\pi^{2}\!\!\! \sum_{mntt'}
\!\!Q^{-1}_{jm\tau t}
	\langle \phi_{mt}\phi_{nt'} \rangle Q^{-1}_{nkt'\tau'} ,
\label{eq:lrci}
\end{equation}
where the correlation function of the dual variables $\phi$ is now to be
calculated using the sine-Gordon-like action \cite{kn:hkle}
\begin{equation}
	S[\phi] =\pi^{2}\sum_{ij}\sum_{tt'}\phi_{it}Q^{-1}_{ijtt'}\phi_{jt'}-
	H\sum_{it}\cos(2\pi\phi_{it})
\label{eq:cos}
\end{equation}
We will calculate the charge-charge correlation function in
a self consistent harmonic approximation which is valid not too close to
the transition point. It amounts to the
replacement of the nonlinear cosine term in
the Hamiltonian by a mass term (or inverse correlation lenght) which is
determined selfconsistently by means of the Bogoliubov variational
principle \cite{kn:SCHA}.
The constant $H$ is related to the fugacity for charges in the original model,
it is $H^{-1}=2\epsilon E_{C}$.
We take a trial action with the $\cos(2\pi\phi_{it})$ replaced by a mass term
\begin{equation}
	S[\phi] =\frac{1}{2}\sum_{ij}\sum_{tt'}\phi_{it}
	[2\pi^{2}Q^{-1}_{ijtt'}+\delta_{ij}\delta_{tt'}\xi^{-2}]\phi_{jt'}
\label{eq:phi}
\end{equation}
and determine the correlation length $\xi$ (or inverse mass) from
the selfconsistency equation
\begin{equation}
	\frac{\xi^{-2}}{4\pi^{2}H}=\exp(-2\pi^{2}\langle
	\phi^{2}\rangle_{\xi^{2}})
\label{eq:frac}
\end{equation}
in the usual way \cite{kn:scha}.
The result for the correlation length is
\begin{equation}
	\xi^{2}= \frac{\epsilon E_{C}}{2 \pi^{3}}\left(\pi
	e^{-\epsilon E_{C} c(\epsilon)/\pi} \right)^{\frac{1}{1-\alpha}} \!\!,
	\;\; \alpha=\frac{\epsilon
E_{C}}{\pi}\left(1+\frac{2}{\epsilon^{2}\omega^{2}_{p}}\right)
\label{eq:xikw}
\end{equation}
and the function $c(\epsilon)$ is of order one. The phase transition is at
$\alpha=1$, which corresponds to $E_{J}/E_{C}=1/\pi^{2}$. Thus, without
vortex fluctuations the phase transition is at a smaller $E_{J}/E_{C}$ value
than
the $2/\pi^{2}$ that follows from a duality argument \cite{kn:sage}.

The correlation function is modified to
\begin{equation}
	\langle q q \rangle_{k,\omega_{\mu}}=
	\frac{k^{2}E_{J}/\epsilon}{\omega^{2}_{\mu}+\tilde{\omega}^{2}_{k}} ,
	\;\;\tilde{\omega}^{2}_{k}= \omega^{2}_{k} + 4\pi^{2} E_{J}\xi^{2}
k^{2}/\epsilon.
\label{eq:aha}
\end{equation}
Thus the spinwave dispersion is affected at small distances (large
$k$-vectors). This hardening of the
spinwave dispersion close to the transition may be interpreted as resulting
from the discreteness
of the Cooper pairs, which makes fluctuations of phase and charge on short
distances unfavourable.
It leads to a mass
\begin{equation}
	M_{v}=\frac{\epsilon}{8\pi\xi^{2}}\ln\left[ 1+\frac{2\pi^{3}\xi^{2}}{\epsilon
E_{C}}\right]
\label{eq:mmm}
\end{equation}
In the limit of small $\xi$
the Eckern-Schmid mass is recovered. An extrapolation to the S-I transition
where $\xi \rightarrow \infty$ yields a mass that vanishes at the transition,
see Fig.2. We find a similar result for the instanton action
\begin{equation}
	S_{inst}=
\frac{\epsilon\omega_{p}}{16\pi\xi^{2}}\left(\sqrt{1+\frac{2\pi^{3}\xi^{2}}
	{\epsilon E_{C}}} -1 \right)
\label{eq:sss}
\end{equation}
Again, the classical result is recovered in the limit $\xi \rightarrow \infty$,
whereas an
extrapolation to the transition gives an instanton action that vanishes. From
the WKB relation between mass, instanton action and barrier height $U_{bar}$
(as discussed in Sect.\ref{sec-cla}), we find that the depinning
current $I_{dep}\sim S^{2}_{inst}/M_{v} \sim 1/\ln(\xi)$ and thus vanishes
algebraically close to
the transition, i.e. $I_{dep}\sim (1-\alpha)$, where $\alpha$ was defined in
Eq.(\ref{eq:xikw}).

With the charge-charge correlation given in Eq.(\ref{eq:aha}) we may calculate
the
spinwave damping of vortex motion due to the coupling to spinwaves beyond the
classical limit. Replacing $\omega_{k}$ by $\tilde{\omega}_{k}$ in
Eq.(\ref{eq:reom}),
the overlap integral over the delta functions only contributes for vortex
velocities
that are higher than a threshold velocity $v_{t}=2\pi\xi\sqrt{E_{J}/\epsilon}$
(see Fig.2). Note that this threshold velocity is independent of the momentum
integration cut-off that is used. Thus, for vortex velocities $v\le v_{t}$
there is
no constant velocity solution to the equation of motion, unless the external
driving current
$I=0$.
Taking into account quantum effects changes the spinwave spectrum in such a way
that the velocity window in which vortices move over the lattice potential
without emitting spinwaves grows larger.
The resulting relation between applied current and vortex velocity is shown in
Fig.1 for
several values of $v_{t}$.
An extrapolation to the S-I transition yields a diverging threshold velocity
and vortices and spinwaves decouple.

\section{Short range Coulomb interactions}
\label{sec-cog}

In this section we consider the case of short range Coulomb interaction $U$,
i.e. the junction capacitance is negligible compared with the capacitance
to the ground. This limit is of more relevance for 2-dimensional
superconducting films.
The system undergoes a $T=0$ phase transition which belongs to
the 2+1-dimensional XY universality class \cite{kn:doni}.
The critical properties are well captured by a
coarse-grained Ginzburg-Landau free energy. The effective free energy
has been derived from the Hamiltonian (\ref{eq:hhh})
using a Hubbard-Stratonovich transformation \cite{kn:doni,kn:cbru}.

It is therefore natural to express the charge-charge correlation function
in terms of a Ginzburg-Landau coarse grained order parameter field.
The charge-charge correlation can be expressed as a functional
derivative of an appropriate generating functional
as follows (see Apppendix \ref{ss-lr} for more details)
$$
	\langle q_{j\tau}q_{k\tau'}\rangle=\frac{\delta^{2}}
	{\delta \mu_{j\tau}\delta \mu_{k\tau'}}
	\ln \left[\int {\cal D}\bar{\psi}{\cal D}\psi\;
	\exp(-F[\bar{\psi},\psi,\mu]) \right],
$$
where
\begin{equation}
	F\!=\!\!\int\!\! d^{3}x \Big\{ \frac{1}{4}\mid\! \vec{\nabla}\psi\!
\mid^{2}\!\!
	+ r\mid\! \psi\! \mid^{2}\!\! + u\mid\!\psi\!\mid^{4}\!\!+
	\zeta\mid\! (\partial_{\tau}-\mu)\psi\! \mid^{2}\Big\} \; ,
\label{eq:free}
\end{equation}
and the coefficients are $r=1/2E_{J}-1/2E_{o}$, $u=7/128E^{3}_{o}$, and
$\zeta=1/32E^{3}_{o}$.
The charge-charge correlation is related to the response of the
system to a twist of the boundary conditions in the time direction.
Its Fourier transform is
\begin{equation}
		\langle q q \rangle_{k,\omega_{\mu}}=
		\zeta\left[2 \langle\psi(\vec{r},\tau) \psi^{*}((\vec{r},\tau)
		\rangle - 4\zeta \!\!\int\!\! d^{2}r d\tau \langle J^{\tau}
		(\vec{r},\tau) J^{\tau}(\vec{0},0)\rangle
		e^{ikr+i\omega\tau} \right]
\label{eq:proj}
\end{equation}
Where the current in the $\tau$ direction $J^{\tau}$ is defined as
\begin{equation}
	J^{\tau}=\frac{1}{2i}
\left\{\psi^{*}(\vec{r},\tau)\partial_{\tau}\psi(\vec{r},\tau)
	- \psi(\vec{r},\tau)\partial_{\tau}\psi^{*}(\vec{r},\tau) \right\}
\label{eq:yo!}
\end{equation}
This type of correlation function was extensively investigated in
\cite{kn:anne}. Due to the isotropy of the model
in space-time the $k=\omega=0$ term in Eq.(\ref{eq:proj}) is proportional
to the superfluid density $\rho_{s}$ of the system.

As discussed previously the adiabatic mass is related to the
zero frequency component of the charge-charge correlation function.
Therefore one obtains
\begin{equation}
	M_{v} \sim \rho_{s} \ln(L/a) + \mbox{ terms not divergent in L} \; ,
\label{eq:mvmv}
\end{equation}
where the terms independent of $L$, that do not diverge with the system size,
may arise from the $k$-dependence of the zero
frequency component of the charge-charge correlation function, however, the
$\ln(L/a)$ is dominant
in the superconducting case.
Close to the transition $\rho_{s}\sim (E_{J}/E_{o}-1)^{\beta}$, with $\beta
\approx 2/3$.
We stress that this result is independent on the particular approximation
we may choose to evaluate the correlation functions.

More care it is needed right at the transition where it can be shown that
the vortex mass {\it does not} vanish; we remind
that it is determined by an integral of the charge-charge correlation
function over the first Brillouin zone.
Although
$\langle q q \rangle_{k=\omega=0}$ vanishes, there is an important
contribution from the k dependence of the correlation function.
Following \cite{kn:anne} it may be calculated employing a 1/N expansion and
to leading order
\begin{equation}
	\langle qq\rangle_{k,\omega_{\mu}} =
	\frac{1}{32\epsilon E_{o}}\sqrt{k^{2}+
	\omega^{2}_{\mu}/4E^{2}_{o}}.
\label{eq:ddd}
\end{equation}
The k-dependence at zero frequency will regularize the k-integral for the mass
and,
as a result it will not vanish at the transition but
becomes independent of the system size
\begin{equation}
	M_{trans} = \frac{\pi^{3/2}}{32E_{o}} \approx 0.44 M_{o}
\label{eq:nec}
\end{equation}
In the superconducting phase the charge-charge correlation function
can be approximated to
\begin{equation}
	\langle qq\rangle_{k,\omega_{\mu}} \sim
	\rho_{s} k^{2}/(k^{2}+4\zeta\omega^{2}_{\mu})
\label{eq:srqq}
\end{equation}
Spinwave damping may be calculated from the equation of motion.
 From Eqs.(\ref{eq:reom}) and (\ref{eq:srqq}) a threshold velocity
$v_{t}=\bar{\omega}_{p}\sqrt{E_{o}/8E_{j}}$ is found,
which for the short range Coulomb interacting does not diverge at the
transition.

We checked the main conclusions of this section performing simulations
that are presented in the next section.

\section{Monte Carlo simulations}
\label{sec-mcs}

We now turn to the Monte Carlo results that allow for a check of the previous
calculations and provide information for the region where the self consistent
harmonic
approximation for the long range case is not valid.
Since the CCVG described by Eq.(\ref{eq:axi}) contains an imaginary coupling
and
long range interactions, it is more convenient \cite{kn:self} to simulate
the system in the equivalent current loop representation Eq.(\ref{eq:cancer}).
The correspondence is simply
$\langle qq\rangle =\langle J^{0}J^{0}\rangle$. The condition that the currents
$J^{\mu}$ be
divergenceless is taken into account by making Monte Carlo steps that preserve
the property
$\nabla_{\mu}J^{\mu}=0$. Thus, we create or annihilate small current loops,
as well as 'periodic current loops' that go through the whole system which is
taken to
have periodic boundary conditions in all three directions. In the case of long
range Coulomb
interaction periodic current loops in the time direction are forbidden since
they violate
charge-neutrality as demanded by the logarithmic Coulomb interaction.

The size in the time direction was taken to be 8, which
corresponds to a temperature $T$ equal to one eight of the plasma frequency,
which is quite low for a JJA.
The simulations were done on lattices of linear dimension varying from 4 to 12
and
the standard Metropolis algorithm was used, with typically 5000 sweeps
through the lattice for equilibration and the same amount for measurement. In
all cases the
full charge-charge correlation function was measured and with the help of
Eq.(\ref{eq:mfc})
the mass was determined. Reliable results for the correlation length would
require
larger systems than we were able to simulate.

For logarithmic Coulomb interaction data for the vortex mass is shown in Fig.3.
The instanton action $S_{inst}$ for a vortex hop from a plaquette to a
neighboring one
and therefore the depinning current behave in a similar way. The critical point
is at
$E_{J}/E_{C}\approx .6$, which agrees with experimental findings \cite{kn:SI}.
Close to the
S-I transition the vortex mass depends strongly on the ratio $E_{J}/E_{C}$.
Note that apart
from small corrections the mass is system size independent.

The results for the vortex mass for onsite Coulomb inter\-action are shown in
Fig.4.
The critical point is at $E_{J}/E_{o}\approx .85$. The collapse of the curves
for different
system sizes (see Fig.4b) demonstrates that the vortex
mass indeed scales with $\ln(L)$ in the SC phase, whereas in the insulating
phase it scales
to zero (see the inset of Fig.4b). If the logarithm of the system size is not
scaled out,
the curves for the mass
in systems of different size approximately cross at the transition
between 0.4 and 0.5 times $M_{o}$,
which is in good agreement with $M_{trans}/M_{o}\approx 0.44$.
The region were the mass is strongly dependent on the ratio of the couplings
is somewhat larger than in the long range case.

\section{conclusion}

We have investigated vortex motion, mass and spinwave damping
in Josephson junction arrays. We first derived a one-vortex effective action
from
which it became clear that dynamical vortex properties are governed by
the charge-charge correlation function. We showed how to recover all known
results
for classical arrays. The Eckern-Schmid mass being proportional
to the junction capacitance, was found to be correct in the limit $E_{J}\gg
E_{C}$,
but also the ground capacitance contributes to the classical mass, which
becomes
system-size dependent.

In the generic situation vortex motion is affected also by the presence
of other vortices (or more precise dipoles in the superconducting state)
which are present because of thermal or quantum (this is the case we
considered in detail) fluctuations. In the quantum regime
close to the S-I transition we investigated the vortex properties
analytically and by means of Monte Carlo simulations.

We considered specifically the two extreme cases in which either the
self-capacitance $C_{o}$
or the junction capaciatnce $C$ was set two zero.
In the case of long range Coulomb interactions ($C_{o}$ equal to zero)
the main conclusions are that the mass and depinning current vanish at the
phase transition
in a way that reflects the nature of the S-I transition, whereas the velocity
window in
which vortices can move without exciting spinwaves grows. Our predictions for
the depinning current are in qualitative agreement with experiment
\cite{kn:herp}.
Our results suggest that ballistic vortex motion may be seen best in arrays
that are close
to the S-I transition.

The limit of on-site Coulomb interaction seems more appropriate for granular
or uniform films. In this case the mass has a logarithmic dependence on the
size of the sample in the superconducting region. It is proportional to
the superfluid density, and therefore shows critical behaviour approaching the
superconductor-insulator transition.
At the transition it does not vanish, but becomes scale independent.

Using Monte Carlo simulations for determining the charge-charge correlation
function
numerically, we were able to verify several conclusions of the self consistent
harmonic approximation (for long range Coulomb interactions) and the
coarse-graining
approach (for short range Coulomb interactions).

In all the results we presented we were mainly concerned with the
superconducting side of the S-I transition, allthough the equations derived
in section \ref{sec-eff} are in principle valid throughout the phase diagram.
We might also investigate the dynamical properties of the vortices
in the insulating region. This is, however, not useful since the vortices are
massless, delocalized, and strongly fluctuating in the insulating phase.
In particular when $E_{C}\gg E_{J}$ the charge on the superconducting
islands is a good quantum number and it is more useful to describe the quantum
dynamics of charges in the resistive (or insulating) region, which may be
investigated using the same techniques as presented here for vortices.
In the resistive (high temperature) phase the concept of vortex is also
not usefull as the system is not globally superconducting and
the superconducting phases of the islands are disordered. We may also
add that the main approximation that leads to the vortex effective action,
namely the truncation of the cumulant expansion to the terms quadratic in
the velocities
may not be justified in the disordered (both resitive and insulating) phases.
Probably in this case a more complicated equation of motion that includes
terms proportional to higher powers of the velocity should be included.

\vspace{.6cm}
\noindent
{\bf Acknowledgements} We thank Gerd Sch\"{o}n, Herre van der Zant, U. Eckern,
W. Elion, G. Falci, A. Groshev and J.E. Mooij for valuable discussions.
R.F and A.v.O. acknowledge a NATO Linkage Grant which enabled the
collaboration.
The work was partially supported by NSF Grant PHY 89-04035 (R.F.) and
'Sonderforschungsbereich 195' (A.v.O.).

\appendix

\section{Duality transformations}
\label{app-dtr}

This appendix reviews shortly how to pass from Eq.(\ref{eq:hhh}) to
Eq.(\ref{eq:axi}).
For more details and the treatment of the external current contribution we
refer to
Ref.\cite{kn:fgs}.
We start from the basic expression for the partition function $Z$
\begin{equation}
	Z=\mbox{Tr}\; \exp(-\beta \hat{H}),
\end{equation}
where $\beta$ is the inverse temperature $T$ and $\hat{H}$ the Hamiltonian
Eq.(\ref{eq:hhh}).
We go over to a Euclidean path-integral formulation by introducing time-slices,
i.e.
dividing $\beta$ in $N_{\tau}$
intervals of size $\epsilon$, such that $N_{\tau}\epsilon=\beta$. Inserting
complete sets
of states at each time slice we arrive at
\begin{equation}
	Z=\sum_{\{q_{i\tau}\}}\int {\cal D}\phi_{i\tau} \exp\left\{-\frac{2\epsilon
E_{C}}
	{\pi}\sum_{ij\tau}
	q_{i\tau}U_{ij}q_{j\tau}+i\sum_{i\tau}q_{i\tau}\dot{\phi}_{i\tau}+
	\epsilon E_{J}\sum_{<ij>\tau}\cos(\phi_{i\tau}-\phi_{j\tau})\right\}
\label{eq:krebs}
\end{equation}
In Fourier components, the electrostatic interaction between
the charges is $U_{k}=2\pi/(k^{2}+\lambda^{2})$,
where the inverse range of the interaction is $\lambda=\sqrt{C_{o}/C}$. It is
related to
the inverse capacitance matrix by $U_{ij}=2\pi C C^{-1}_{ij}$.
Now we make the Villain approximation \cite{kn:vill} for the cosine term
\begin{equation}
	\exp\left(\epsilon E_{J}\cos(\phi_{i\tau}-\phi_{j\tau})\epsilon
E_{J}\right)\approx
	\sum_{\{n_{ij,\tau}\}} \exp\left(-\frac{\epsilon E_{J}}{2}f(\epsilon E_{J})
	(\phi_{i\tau}-\phi_{j\tau}-2\pi n_{ij,\tau} )^{2}\right),
\end{equation}
where $n$ is a directed discrete field that lives on the bonds between lattice
sites. The function
$f$ equals unity if its argument is large, i.e. when the product
$\epsilon E_{J}$ is not too small. After a subsequent Poisson resummation, i.e.
writing
\begin{equation}
	\sum_{\{n_{ij,\tau}\}}F[n_{ij,\tau}]=
	\sum_{\{J_{ij,\tau}\}}\int {\cal D}n F[n_{ij,\tau}]
	\exp(2\pi i \sum_{ij,\tau} n_{ij,\tau}J_{ij,\tau}) ,
\label{eq:fish}
\end{equation}
an integration over the fields $n_{ij,\tau}$ and the phases $\phi_{i\tau}$
yields
a representation in terms of divergenceless discrete current loops
\begin{equation}
	Z=\sum_{\{J^{\mu}_{i,\tau}\}} \delta(\nabla_{\mu}J^{\mu})
	\exp\left\{-\frac{2\epsilon E_{C}}{\pi}\sum_{ij,\tau}J^{0}_{i,\tau}
	U_{ij}J^{0}_{j,\tau}-\frac{1}{2\epsilon E_{J}}
	\sum_{i,\tau,a}(J^{a}_{i,\tau})^{2}\right\},
\label{eq:cancer}
\end{equation}
where $a=x,y$ and $\mu=x,y,0$.
Here the time components $J^{0}$ of the current are simply the charges. The
vortex degrees
of freedom may now be extracted by solving the constraint in
Eq.(\ref{eq:cancer}) by
writing $J^{a}_{i\tau}=\epsilon^{ab}\nabla_{b}\psi_{i\tau}- O^{a}q_{i\tau}$
(the operator $O^{a}$ denotes the line integral in direction $a$, i.e. it is
the inverse of $\nabla_{a}$) and making a final
Poisson resummation on the discrete field $\psi$. Making use of the identity
$\Theta_{ij}=\epsilon_{ab}\nabla^{a}O^{b}G_{ij}$ Eq.(\ref{eq:axi}) results.

Note that we did not keep track of determinants, as they are irrelevant for the
present
purpose. The requirements that both the factorization of $\exp(-\beta H)$ and
the Villain approximation are valid restricts $\epsilon$ to be of
the order of the inverse of the plasma frequency
$\omega_{p}=\sqrt{8E_{J}E_{C}}$ for long
and $\bar{\omega}_{p}=\sqrt{8E_{J}E_{o}}$ for short range Coulomb interactions,
i.e.
we take $\epsilon=p/\omega_{p}$ or $\epsilon=p/\bar{\omega}_{p}$. Details and
numerical
factors may depend on the exact choice for $p$. We take $p=1$, except for
the self consistent harmonic approximation where we take $p=\sqrt{2}$.
The plasma frequency is the natural frequency for spinwaves.

\section{The charge-charge correlation function}
\label{app-ccc}

In general the charge-charge correlation function may be expressed
as a functional derivative of the free energy in the following way
\begin{equation}
	\langle q_{jt}q_{kt'}\rangle=\frac{\delta^{2}}
	{\delta \mu_{jt}\delta \mu_{kt'}}\ln\left[
	\int {\cal D}\xi \exp\left(-S[\xi]+\sum_{i\tau} \mu_{i\tau} q_{i\tau}\right)
	\right]_{\mu=0}.
\label{eq:Paula}
\end{equation}
Here $\xi$ denotes the field or fields that are integrated (or summed) over,
$S[\xi]$ a corresponding action and $\int {\cal D}\xi$ the appropriate measure.

\subsection{long range Coulomb interactions}
\label{ss-lr}

For instance, if one takes the action to be the charge part of the CCVG action,
as is done in section \ref{sec-sca}, one has
\begin{equation}
	\langle q_{kt}q_{lt'}\rangle=\frac{\delta^{2}}{\delta \mu{kt}\delta
\mu_{lt'}}\ln\left[
	\sum_{q_{i\tau}}
\exp\left(-\sum_{ij\tau\tau'}q_{i\tau}Q_{ij\tau\tau'}q_{j\tau'}+
	\sum_{i\tau} \mu_{i\tau} q_{i\tau}\right)
	\right]_{\mu=0}.
\label{eq:qqq}
\end{equation}
The kernel $Q$ was defined in Eq.(\ref{eq:chke}).
Now the partition function may be rewritten in terms of different (dual) fields
and if
we keep track of the 'currents' $\mu$ during the transformation
we may express the charge-charge correlation function
in terms of the correlation function of the new fields.
Applying this strategy to Eq.(\ref{eq:qqq}), we find after a Poisson
resummation
\begin{eqnarray}
\nonumber
	\langle q_{kt}q_{lt'}\rangle=\frac{\delta^{2}}{\delta \mu_{kt}\delta
\mu_{lt'}}\ln{\Big [}
	\sum_{l_{i\tau}} \exp{\Big (}-\pi^{2}\sum_{ij\tau\tau'}l_{i\tau}
	Q^{-1}_{ij\tau\tau'}l_{j\tau'}+\\
	+i\pi\sum_{ij\tau\tau'}l_{i\tau}Q^{-1}_{ij\tau\tau'}\mu_{j\tau'}
	+\frac{1}{4}\sum_{ij\tau\tau'}\mu_{i\tau}Q^{-1}_{ij\tau\tau'}\mu_{j\tau'}{\Big
)}
	{\Big ]}_{\mu=0} ,
\label{eq:qql}
\end{eqnarray}
from which we read of immediately that
\begin{equation}
	\langle q_{kt}q_{lt'}\rangle=
	\frac{1}{2} Q^{-1}_{kltt'} -\pi^{2}\! \sum_{mn\tau\tau'} \!\!Q^{-1}_{kmt\tau}
	\langle l_{m\tau}l_{n\tau'} \rangle Q^{-1}_{nl\tau' t'} .
\label{eq:qqp}
\end{equation}
This has simplified the problem of calculating the correlation function
considerably, since
the new field $l$ interacts with the kernel $Q^{-1}$ which is,
in contrast to the original kernel $Q$, short range in both the space and time
directions.
This  discrete Gaussian model is therefore convenient for Monte Carlo
simulations.

Another representation for the correlation function is found by applying an
inverse Villain
approximation, see chapter 11 of Ref.\cite{kn:hkle} for details.
This enables one to rewrite the
partition function as a path integral over a continuous field with sine-Gordon
action. The
result is very similar to Eq.(\ref{eq:qqp})
\begin{equation}
	\langle q_{j\tau}q_{k\tau'}\rangle
	=\frac{1}{2} Q^{-1}_{jk\tau\tau'} -\pi^{2}\!\!\! \sum_{mntt'}
\!\!Q^{-1}_{jm\tau t}
	\langle \phi_{mt}\phi_{nt'} \rangle Q^{-1}_{nkt'\tau'} ,
\end{equation}
but now the correlation function of the dual variables $\phi$ is now to be
calculated using the action \cite{kn:hkle}
\begin{equation}
	S[\phi] =\pi^{2}\sum_{ij}\sum_{tt'}\phi_{it}Q^{-1}_{ijtt'}\phi_{jt'}-
	H\sum_{it}\cos(2\pi\phi_{it})
\end{equation}
In this formulation we may employ the self consistent harmonic approximation
\cite{kn:SCHA}.

\subsection{short range Coulomb interactions}
\label{ss-sr}

Here we outline the derivation of Eq.(\ref{eq:free}) in the text. For details
we refer to
Ref.\cite{kn:cbru}. In the case of short range interaction, the critical
properties of the
system are well described by a coarse grained free energy.
The generating functional that we consider is the partition function
Eq.(\ref{eq:krebs}) with
a coupling to a current $\mu$ as in Eq.(\ref{eq:Paula}). After performing the
integration over
${\cal D}q$ we have in terms of phase-variables only
\begin{eqnarray}
\nonumber
	\langle q_{jt}q_{kt'}\rangle=\frac{\delta^{2}}{\delta
	\mu_{jt}\delta \mu_{kt'}}  \ln{\Big [}\int {\cal D}\phi \exp{\Big (}
	\frac{1}{8e^{2}} \sum_{ij}\int d\tau
\left[\dot{\phi}_{i}(\tau)+i\mu_{i}(\tau)\right]
	C_{ij} \left[\dot{\phi}_{j}(\tau)+i\mu_{j}(\tau)\right]-\\
	-E_{J} \sum_{<ij>} \int d\tau \cos(\phi_{i}-\phi_{j})  {\Big )}{\Big ]} \; ,
\label{eq:blabla}
\end{eqnarray}
where the measure ${\cal D}\phi$ still contains a summation over winding
numbers $m_{i}$
(i.e. $\phi_{i}(\beta)=\phi_{i}(0)+2\pi m_{i}$) in order to account for the
discreteness of the
charges.
Decoupling the Josephson term by means of a Hubbard-Stratonovich
transformation one gets:
\begin{eqnarray}
\nonumber
	\langle q_{jt}q_{kt'}\rangle=\frac{\delta^{2}}{\delta \mu_{jt}\delta
\mu_{kt'}}  \ln{\Big [}
	\int {\cal D}\psi {\cal D}\bar{\psi} \exp \left\{\frac{1}{2E_{J}} \sum_{ij}
	\int d\tau \left[\bar{\psi}_{i}(\tau) t^{-1}_{<ij>}\psi_{j}(\tau) \right]
\right\}\\
	{\Big \langle} \exp\left\{\sum_{i}\int d\tau\left[\bar{\psi}_{i}(\tau)
	e^{i\phi_{i}(\tau)}\right]\right\} {\Big \rangle} {\Big ]} \; ,
\label{eq:nose}
\end{eqnarray}
where the matrix $t_{<ij>}$ is one for nearest neighbors and zero otherwise.
The average in the
last factor is with respect to the remainder of the action in
(\ref{eq:blabla}).
Close to the transition point we may expand the average
in powers (cumulants) of the fields $\psi$; this expansion yields the
generating functional
used in the paper. We note that the currents $\mu$ enter the time derivatives
in the generating
functional Eq.(\ref{eq:free}) in a gauge invariant way.

\vspace{1cm}

\begin{large}
\noindent
Figure Captions:
\end{large}

\vspace{1cm}

\noindent
Fig. 1 : Driving-current vs. vortex velocity
relations for long range Coulomb interactions and several couplings.
 From left to right: $v_{t}=0$ (for $E_{J}/E_{C}\gg 1$,
the classical limit, see \cite{kn:ulig}), $v_{t}/\omega_{p}=0.5$ and
$v_{t}/\omega_{p}=1$.
The vertical dotted line indicates the minimum velocity a vortex
needs to move ballistically over the pinning potential.

\vspace{1cm}

\noindent
Fig. 2 : Mass, threshold velocity and coherence length as from the self
consistent
harmonic approximation.

\vspace{1cm}

\noindent
Fig. 3 : Monte Carlo result for the mass with long range interactions.
Shown are a 6x6x8 (crosses) and a 10x10x8 (diamonds) system.
The statistical errors are of the order of the symbol size.

\vspace{1cm}

\noindent
Fig. 4 :  Monte Carlo results for the mass with on-site interactions.
a) Shown are a 4x4x8 (lower curve), 6x6x8, 8x8x8, 10x10x8 and a 12x12x8 (upper
curve) system.
The statistical errors are of the order of the symbol size.

b) The mass for on-site interactions with the logarithmic of the system
size scaled out. Clearly the curves collapse on one line in the SC phase, but
{\it not} in the critical region. Inset: The size dependence of the mass with
the
logarithm scaled out for $E_{J}/E_{o}=$1.45, 1.1, .95, .85, .75 (from top to
bottom).

\end{document}